\documentclass[RNAAS]{aastex62}
\usepackage{amsmath}
\usepackage{amsfonts}

\shorttitle{PBH}
\shortauthors{Lu et al.}

\newcommand{\msun}{\ensuremath{M_{\odot}}}

\begin{document}

\title{Primordial Black Hole Microlensing: The Einstein Crossing Time Distribution}

\author[0000-0001-9611-0009]{Jessica R. Lu}
\affiliation{Department of Astronomy, University of California, Berkeley, CA, USA 94720}

\author{Casey Y. Lam}
\affiliation{Department of Astronomy, University of California, Berkeley, CA, USA 94720}

\author[0000-0002-7226-0659]{Michael Medford}
\affiliation{Department of Astronomy, University of California, Berkeley, CA, USA 94720}
\affiliation{Lawrence Berkeley National Laboratory, 1 Cyclotron Rd, Berkeley, CA 94720}
\affiliation{Lawrence Livermore National Laboratory, 7000 East Ave, Livermore, CA 94550}

\author[0000-0003-0248-6123]{William Dawson}
\affiliation{Lawrence Livermore National Laboratory, 7000 East Ave, Livermore, CA 94550}

\author[0000-0003-2632-572X]{Nathan Golovich}
\affiliation{Lawrence Livermore National Laboratory, 7000 East Ave, Livermore, CA 94550}


\section{Introduction}

Gravitational microlensing is one of the few means of finding primordial black holes (PBHs), if they exist. 
Recent LIGO detections of 30 $\msun$ black holes have re-invigorated the search for PBHs in the 10-100 $\msun$ mass regime. 
Unfortunately, individual PBH microlensing events cannot easily be distinguished from stellar lensing events from photometry alone. 
However, the distribution of microlensing timescales  ($t_E$, the Einstein radius crossing time) can be analyzed in a statistical sense using models of the Milky Way with and without PBHs. While previous works have presented both theoretical models and observational constrains for PBHs \citep[e.g.][]{Calcino:2018,Niikura:2019}, surprisingly, they rarely show the observed quantity -- the $t_E$ distribution -- for different abundances of PBHs relative to the total dark matter mass ($f_{PBH})$.

\section{Model for Primordial Black Hole Lenses}

We present a simple calculation of how the $t_E$ distribution changes between models with and without PBHs. 
We utilize PopSyCLE \citep{Lam:2019} to simulate microlensing events for a $3.74$ deg$^2$ field towards the Galactic bulge that includes stars and stellar-mass compact objects, but not PBHs.  PopSyCLE output is then modified to add PBH lensing events via the  procedure below. Several simplistic assumptions are made when injecting PBHs:
\begin{itemize}
    \setlength\itemsep{-0.05in}
    \item The spatial and velocity distribution of PBHs follows the stellar halo.
    \item The mass distribution of PBHs is Gaussian with $\overline{m}_{PBH} = 30 \msun$ with a spread of $\sigma_{M_{PBH}} = 20$\msun. 
    \item The total mass of the dark matter halo (including PBHs) is $M_{DM} = 10^{12} \msun$. 
    \item The total mass of the stellar halo is $M_{H\star} = 10^9 \msun$.
\end{itemize}
Microlensing rates depend on the number of lens objects. On average, the total number of primordial black holes in the Milky Way, $N_{PBH}$, is 
\begin{align}
    N_{PBH} = \frac{f_{PBH} M_{DM}}{\overline{m}_{PBH}}
\end{align}
where $\overline{m}_{PBH}$ is the mean mass of a PBH and $f_{PBH}$ is the fraction of the halo mass in PBHs, which is a free parameter. The number of halo stars in the Milky Way is
\begin{align}
    N_{H\star} = \frac{M_{H\star}}{\overline{m}_{H\star}}.
\end{align}
where $\overline{m}_{H\star}$ is the mean mass of a halo star.
The PopSyCLE synthetic microlensing survey covers a small fraction of the sky towards the Galactic Bulge, in a similar direction as OGLE and MOA on-sky surveys.
Thus we need to convert the all-sky $N_{PBH}$ into the number of PBHs that are lensed in some survey, $N_{PBH,S,L}$.
For a survey, $S$, with a survey duration of $T_S$, the number of lensed objects of any type is \citep{Paczynski:1986},
\begin{align}
    N_{S,L} \approx \frac{2}{\pi} \frac{T_S}{\overline{t}_E}  \tau N_{\star,S}
\end{align}
where $\overline{t}_E$ is the mean Einstein crossing time, $\tau$ is the lensing optical depth, and $N_{\star,S}$ is the number of source stars monitored in the survey typically coming from the bulge, disk, and a small number of stars from the halo. 
Given that PopSyCLE tells us the number of lensed halo stars in our survey, $N_{H\star,S,L}$, we need only consider the ratio of events,
\begin{align}
    \frac{N_{PBH,S,L}}{N_{H\star,S,L}} = 
    \left(\frac{\overline{t}_{E,H\star}}
         {\overline{t}_{E,PBH}}\right) 
    \left(\frac{ \tau_{PBH}}{ \tau_{H\star}}\right).
\end{align}
The Einstein crossing time, $t_E$, is given by $t_E = \theta_E / \mu_{rel}$ where 
\begin{align}
    \theta_E = \sqrt{\frac{4 G m}{c^2} (d_L^{-1} - d_S^{-1})}
\end{align}
is the angular Einstein radius and $\mu_{rel}$ is the source-lens relative proper motion. Assuming that the distance and proper motion distribution is identical for halo stars and PBHs, the $\overline{t}_E$ ratio gives $(\overline{m}_{H\star} / \overline{m}_{PBH})^{1/2}$.
The optical depth ratio is $N_{PBH,S} \theta_{E,PBH}^2 /N_{H\star,S} \theta_{E,H\star}^2$, which is independent of the mass of PBHs and halo stars. Thus, the number of lensed PBHs in the survey becomes
\begin{align}
    N_{PBH,S,L} 
     & = N_{H\star,S,L} 
         \left( \frac{N_{PBH,S}}{N_{H\star,S}} \right)
         \left( \frac{\overline{m}_{PBH}}{\overline{m}_{H\star}} \right)^{1/2} \\
    & = N_{H\star,S,L} 
         \left( \frac{f_{PBH} \cdot M_{DM}}{M_{H\star}} \right)
         \left( \frac{\overline{m}_{H\star}}{\overline{m}_{PBH}} \right)^{1/2}
\end{align}
We note that our simple approximation that the PBH velocity distribution is identical to the halo distribution is only valid when the PBHs make up a small fraction of the halo mass and the gravitational potential is dominated by some other form of dark matter. Thus, we only consider $f_{PBH} \lesssim 0.3$.

We inject the above number of lensed PBHs into the PopSyCLE simulation.
PBHs are injected by randomly drawing from other halo star lensing events and modifying the lens mass, $m_L$, and Einstein crossing time, $t_E$, using
\begin{align}
    m_{L,PBH} &= Norm(\mu=30 \msun, \sigma=20 \msun) \\
    t_{E,PBH} &= t_{E,orig} * \sqrt{m_{L,PBH} / m_{L,H\star}}. \\
\end{align}
We adopt this mass distribution as recent detections of 30 $\msun$ black holes with LIGO have renewed interest in this mass range \citep{Carr:2016}.

\section{The $t_E$ Distribution}
The simulated 3.74 deg$^2$ field of view towards the inner Galactic Bulge contains $1.7 \times 10^9$ stars when no observational cuts are applied. 
Within this field, there are $5\times10^5$ microlensing events in a 1000 day survey before adding PBHs.
PBH lenses contribute an additional $3.8\times10^4$, $1.2\times10^{5}$, and $2.3\times10^5$ events for $f_{PBH}=$0.05, 0.15, and 0.30, respectively.

\begin{figure}
\begin{center}
\includegraphics[scale=0.23]{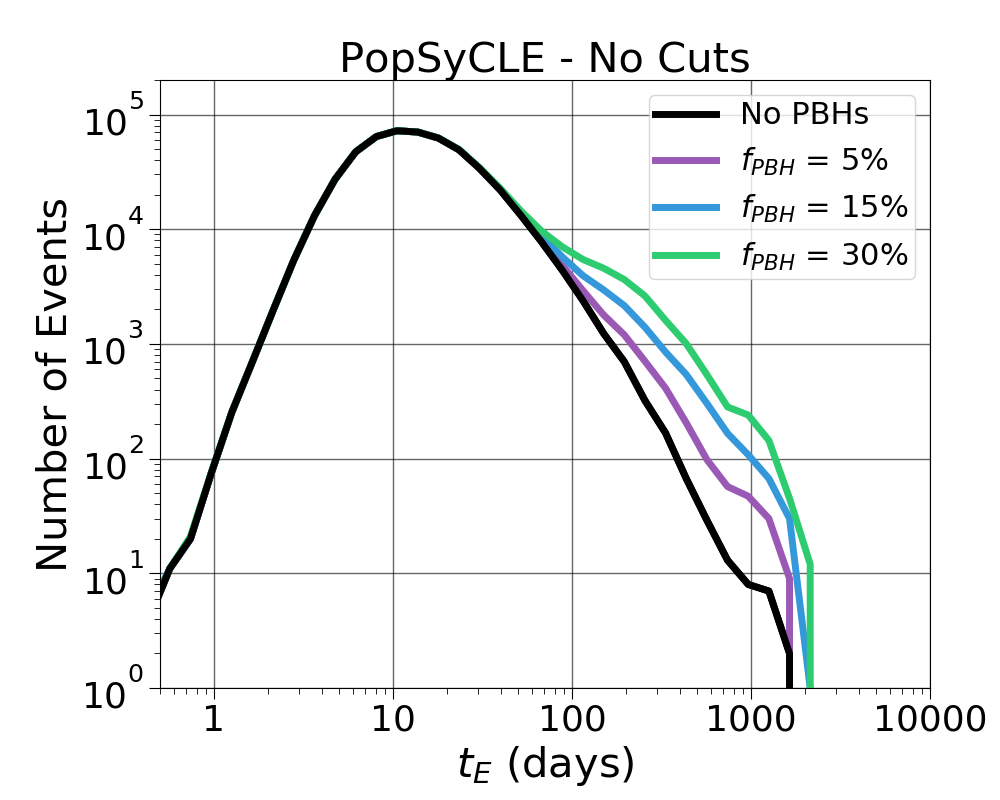}
\includegraphics[scale=0.23]{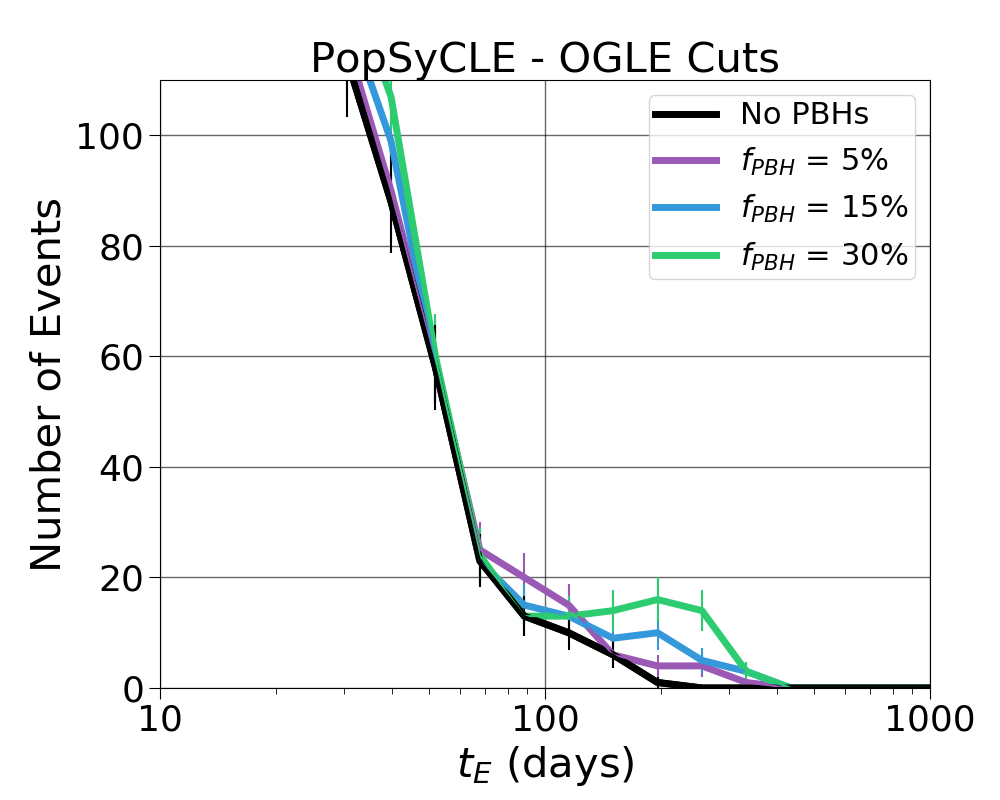}
\includegraphics[scale=0.23]{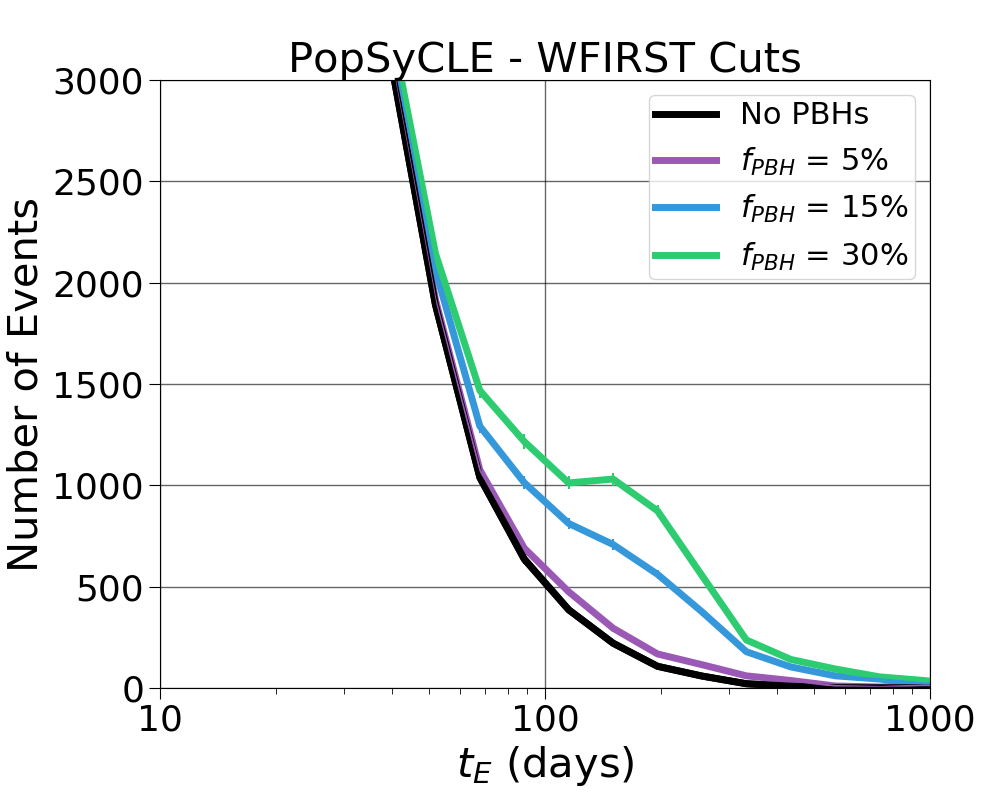}
\end{center}
\caption{The $t_E$ distribution when the Milky Way dark matter halo ($M_{DM}=10^{12} \msun$) is composed of different fractions of primordial black holes (specified by $f_{PBH}$). The PBH masses are assumed to follow a normal distribution peaking at $30 \msun$ with a spread of $20 \msun$.
{\em Left:} No observational cuts. {\em Middle: } OGLE-style observational cuts on a linear scale shows very little sensitivity to PBHs. {\em Right:} WFIRST-style observational cuts enable a strong detection of a PBH signal for $f_{PBH}>0.05$. 
\label{fig:pbh_tE}
}
\end{figure}

The $t_E$ distribution is shown in Figure \ref{fig:pbh_tE}
and is enhanced at long timescales as $f_{PBH}$ increases. 
Also shown is the resulting $t_E$ distribution after observational cuts are applied in a manner suitable for an OGLE \citep{Udalski:2008} or WFIRST \citep{Penny:2019} microlensing survey. 
Cuts are often made on the impact parameter ($u_0$), which is the closest on-sky separation normalized by the Einstein radius, the difference between the baseline and peak magnitude ($\Delta m$),  the source flux fraction ($f_{SFF}$), which is the ratio of unlensed source flux divided by the baseline flux that includes neighboring stars in the beam, and the signal-to-noise at the peak ($SNR_{peak}$).
OGLE observable events include only those with baseline magnitude of I$<$22 mag, $u_0<1$, $\Delta m > 0.1$, and $SNR_{peak}>3$.
WFIRST observable events include only those with baseline magnitude of H$<26$ mag, $u_0<2$, $\Delta m > 0.1$, and $f_{SFF}>0.1$.
The total number of non-PBH microlensing events is reduced, largely due to the magnitude cuts, from $5\times10^5$ to $1.7\times10^2$ and $9.5\times10^4$ for OGLE and WFIRST, respectively. While the PBH signal is difficult to detect in older surveys such as MACHO and OGLE-III, the PBH signal is easily detectable when $f_{PBH}>0.05$ in any multi-year microlensing surveys with $>10^4$ total events, modulo long time-scale systematics that could decrease the sensitivity.

\vspace{0.1in}
\noindent
{\it Acknowledgements:} We thank Alex Drlica-Wagner and George Chapline for useful discussions related to this work. We acknowledge support from the NASA WFIRST SIT Program (NNG16PJ26C), the UCOP and UC Lab Fees Research Program (LGF-19-600357), and the U.S. DoE LLNL (DE-AC52-07NA27344) and LLNL-LDRD Program (17-ERD-120).

\bibliography{pbh_references.bib}

\end{document}